\documentclass[sigconf]{acmart}
\AtBeginDocument{%
  }

\begin{document}

\title{Sustainable Graph Analytics Workload Scheduling with Evolutionary Reinforcement Learning in Edge-Cloud Systems}

\author{Pallavi Ramicetty}
\affiliation{%
  \institution{Colorado State University}
  \city{Fort Collins}
  \country{USA}}
\email{pallavi.ramicetty@colostate.edu}

\author{Hayden Moore}
\affiliation{%
  \institution{Colorado State University}
  \city{Fort Collins}
  \country{USA}}
\email{hayden.moore@colostate.edu}

\author{Sirui Qi}
\affiliation{%
  \institution{Colorado State University}
  \city{Fort Collins}
  \country{USA}}
\email{alex.qi@colostate.edu}

\author{Akhirul Islam}
\affiliation{%
  \institution{IIIT Guwahati}
  \city{Guwahati}
  \country{India}}
\email{akhirul.islam@iiitg.ac.in}

\author{Manojit Ghose}
\affiliation{%
  \institution{IIIT Guwahati}
  \city{Guwahati}
  \country{India}}
\email{manojit@iiitg.ac.in}

\author{Dejan Milojicic}
\affiliation{%
  \institution{Hewlett-Packard Labs}
  \city{Milpitas}
  \country{USA }}
\email{dejan.milojicic@hpe.com}

\author{Cullen Bash}
\affiliation{%
  \institution{Hewlett-Packard Labs}
  \city{Milpitas}
  \country{USA}}
\email{cullen.bash@hpe.com}

\author{Sudeep Pasricha}
\affiliation{%
  \institution{Colorado State University}
  \city{Fort Collins}
  \country{USA}}
\email{sudeep@colostate.edu}

\renewcommand{\shortauthors}{Ramicetty et al.}
\begin{abstract}
Graph analytics powers modern intelligent systems such as smart cities, cyber-physical infrastructure, IoT security, and large-scale social networks. As these workloads scale in complexity, their execution in heterogeneous edge-cloud environments results in higher energy use and carbon emission footprint. To address this challenge, we propose MERSEM, a multi-objective evolutionary reinforcement learning framework for sustainable edge-cloud system management. MERSEM integrates evolutionary search with reinforcement learning (RL) to solve the problem of graph workload allocation and scheduling. The evolutionary component explores diverse global solutions, while the RL agent refines decisions through adaptive local optimization. The framework is designed to jointly minimize service-level agreement (SLA) violations and carbon emissions by considering dynamic carbon intensity, resource heterogeneity, and workload characteristics. Experimental results demonstrate that MERSEM outperforms the state-of-the-art with up to 45$\%$ SLA violation reductions and up to 12$\%$ carbon emission reductions.
\end{abstract}





\settopmatter{printacmref=false}
\renewcommand\footnotetextcopyrightpermission[1]{}
\pagestyle{empty}
\maketitle

\section{Introduction}
Graph analytics and graph neural network (GNN) workloads are foundational to modern intelligent systems, including smart cities, cyber-physical infrastructure, IoT security, and large-scale social networks. These applications rely on graph operations such as traversal, aggregation, and neighborhood sampling, which exhibit irregular execution patterns, data dependencies, and frequent synchronization. As the number of connected IoT devices continues to grow, projected to reach nearly 39 billion by 2030 \cite{sinha2025iot}, the volume and computational complexity of graph workloads generated at the network edge are increasing rapidly.

To meet stringent latency requirements, graph workloads are commonly offloaded from resource-constrained edge devices to nearby fog nodes or centralized cloud datacenters. While edge-cloud execution enables scalable processing of large graphs, it introduces a fundamental trade-off between service-level agreement (SLA) compliance and environmental sustainability. Offloading decisions affect not only communication latency and queueing delays but also the energy consumption and carbon emissions of geographically distributed heterogeneous computing infrastructure. Recent estimates indicate that cloud and edge computing infrastructure account for a significant and growing share of global electricity use and associated carbon emissions \cite{alsharif2025fog}.

Carbon-aware workload management has therefore emerged as a key research direction for sustainable computing systems. Prior studies have demonstrated that intelligent task placement and scheduling can reduce energy use and operational carbon emissions by exploiting spatial and temporal variations in workload characteristics, resource efficiency, and electricity carbon intensity \cite{alsharif2025fog}. However, most existing carbon-aware fog and cloud scheduling frameworks (1) overlook directed acyclic graph (DAG)-induced execution bottlenecks, (2) fail to explicitly enforce application-level SLA constraints, or (3) rely on static or heuristic optimization strategies that struggle to adapt to dynamic system conditions such as fluctuating carbon intensity, workload mix, and resource availability. As a result, jointly optimizing SLA compliance and carbon emissions for graph workloads in heterogeneous edge-cloud systems remains an open challenge.
To address these challenges, we propose MERSEM, a Multi-Objective Evolutionary Reinforcement Learning framework for Sustainable Edge-Cloud Management. MERSEM is designed specifically for graph analytics workloads and co-optimizes SLA violation rate and operational carbon emissions under dynamic conditions. The key contributions of this work are:
\begin{itemize}
\item We formulate a multi-objective graph workload scheduling problem that optimizes SLA compliance and carbon emissions in heterogeneous edge-cloud systems.
\item We propose MERSEM, a novel hybrid evolutionary and reinforcement learning framework that combines global exploration with adaptive local optimization under dynamic system conditions to enable scalable and flexible workload placement across edge, fog, and cloud layers.
\item We develop a comprehensive model capturing graph workload characteristics, platform heterogeneity, latency, energy consumption, and carbon intensity.
\end{itemize}

The remainder of this paper is organized as follows. Section 2 reviews related work. Section 3 presents the system and workload models. Sections 4 and 5 describe the problem formulation and the MERSEM framework. Section 6 evaluates the proposed approach, and Section 7 concludes the paper.
\section{Related Work}
Resource management in fog and edge-cloud environments has been widely studied to improve latency, cost, and resource utilization. Wu et al. proposed a semi-Markov decision process–based offloading strategy to minimize delay in vehicular fog systems \cite{wu2023vehicular}, while Etemadi et al. introduced deep learning–based auto-scaling for dynamic IoT workloads \cite{etemadi2021autoscaling}, and Gupta et al. used Q-learning for UAV-assisted fog offloading \cite{gupta2024uav}. However, these methods do not explicitly consider application-level SLAs or sustainability.

Recent work has focused on energy- and carbon-aware scheduling. Huang et al. used Lyapunov optimization to reduce energy consumption under delay constraints \cite{huang2022lyapunov}. Pournazari et al. proposed carbon-aware metaheuristics for optimizing emissions, energy, and makespan \cite{pournazari2025carbon}, while Moore et al. studied joint carbon and water optimization for LLM inference in cloud systems \cite{moore2025llm}. Chancerel et al. analyzed lifecycle carbon footprints in renewable-powered fog systems \cite{matteo2025lifecycle}, and Islam et al. applied reinforcement learning for energy-harvesting edge offloading \cite{islam2025rl}. Despite progress, most approaches treat workloads as independent tasks and ignore graph-based execution patterns and SLA constraints.

Multi-objective optimization methods such as NSGA-II \cite{deb2002nsga2} and NSGA-III \cite{deb2014nsga3} are widely used for Pareto optimization in distributed systems. Hybrid approaches include game-theoretic optimization for energy and network costs \cite{hogade2022energy} and machine learning–aided evolutionary frameworks for carbon, water, and cost optimization \cite{qi2026shieldeb}. However, these methods are mainly cloud-centric and do not address graph workloads or heterogeneous edge–cloud environments.

In summary, existing approaches either optimize performance or sustainability, but rarely both. Carbon-aware methods often ignore SLAs and graph structures, while hybrid RL-evolutionary techniques have not been applied to carbon-aware graph analytics in heterogeneous environments. MERSEM bridges this gap by introducing a hybrid evolutionary reinforcement learning framework that models DAG-based workloads, SLA constraints, infrastructure heterogeneity, and dynamic carbon intensity.
\section{System Models}
\begin{figure}[!t]
  \centering
  \includegraphics[width=\linewidth]{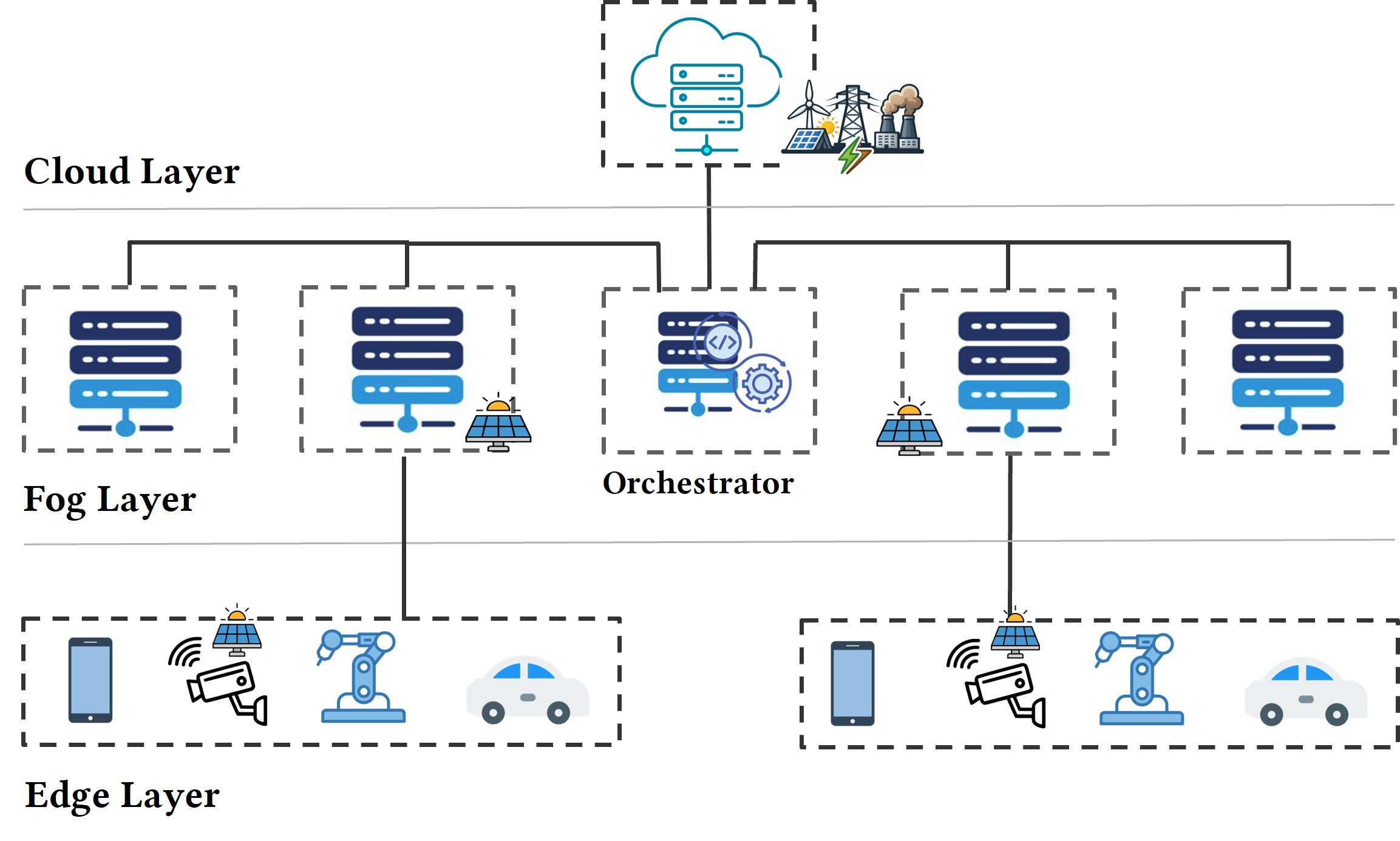}
  \caption{Edge-Fog-Cloud Environment.} \label{fig1:edge-fog-cloud-architeture}
\end{figure}

Our framework targets a multi-tier environment consisting of edge devices, fog datacenters, and a centralized cloud datacenter, as shown in Fig. \ref{fig1:edge-fog-cloud-architeture}. Job requests originate at the edge layer and may be executed locally or offloaded to fog or cloud resources based on system conditions such as resource availability, computation capacity, energy, and sustainability constraints. The system model captures heterogeneous computing resources, location-dependent carbon intensity, and graph workload characteristics (computational and communication requirements).
\subsection{Edge-Fog-Cloud Computing}
We model a heterogeneous edge-fog-cloud architecture that processes latency-sensitive workloads across three layers. 

\textit{1) Edge Layer}: This layer includes edge devices (EDs) such as mobile devices and smart cameras. Mobile devices generate user tasks and are typically powered by grid electricity, although some may use renewable energy sources, resulting in negligible carbon emissions. Smart cameras generate video analytics workloads and may utilize solar energy when available, switching to grid electricity otherwise. Edge devices support local computation.

\textit{2) Fog Layer:} This layer consists of fog datacenters (FDCs) located close to edge devices. Each FDC contains heterogeneous servers that host multiple virtual machines (VMs) with varying computational capacities and power profiles. These facilities are powered by a mixture of renewable energy and grid electricity, resulting in location-dependent carbon emissions.

\textit{3) Cloud Layer}: The cloud datacenter (CDC) provides resources with significantly higher computational and memory capacity than edge or fog resources. It relies on grid electricity, and its carbon emissions reflect the regional energy mix.

A centralized orchestrator manages job execution across the edge-fog-cloud infrastructure. It determines whether jobs generated at edge devices should be executed locally or remotely and selects the destination node accordingly. The objective is to satisfy graph workload SLA requirements while minimizing environmental impact. The orchestrator is assumed to have full knowledge of job characteristics, available resources, and carbon intensity at each node.

Each computing node $n \in N$ is characterized by CPU capacity $C_n$ (MIPS), memory capacity $M_n$ (gigabytes), storage capacity $S_n$ (gigabytes), idle power consumption $P_n^{idle}$ (watts), and maximum power consumption $P_n^{max}$ (watts). Fog and cloud datacenters support concurrent execution of multiple jobs, whereas edge devices execute at most one job at a time due to limited resources.
\subsection{Workload Execution Model}
We use the large-scale graph workflow dataset from \cite{rezaee2020dataset}, which contains about half a million jobs and 1.3 million tasks. Jobs are represented as directed acyclic graphs (DAGs), where nodes represent computational tasks and edges denote precedence dependencies. Each task includes attributes such as millions of instructions (MI), input data size, execution time, memory, and CPU requirements, while job-level properties include the DAG structure, maximum parallel tasks, deadlines, and memory requirements. The system operates in preset discrete time intervals $E$, defined as epochs. During each epoch $e$, a set of jobs $J_e$ is generated by geographically distributed edge devices.

Each job is assigned to a single virtual machine (VM). Multiple jobs may be mapped to the same VM, but they execute sequentially, and a job cannot be partitioned across multiple VMs. A job can be uniquely referenced as $J_{i,j}$ where $i$ is the device index at which it originated, and $j$ is the job index on that device. A job is defined as a DAG in which tasks wait for their predecessors to complete, while independent tasks can be executed in parallel at the same execution level. The degree of parallelism is determined by the DAG structure and the number of available CPU cores on the assigned VM. For communication modeling, we consider transmission delay for the input data of a job.
\subsection{Latency Model}
To evaluate job execution performance in the edge-fog-cloud infrastructure, we model the end-to-end latency experienced by each job. The total latency includes communication delay, queuing delay, and execution time. Jobs generated at edge devices may execute locally or be offloaded to fog or cloud datacenters. When offloading occurs, communication delays arise due to network transmission. Edge devices communicate with fog datacenters through WiFi and metropolitan area networks (MAN), while communication with the cloud involves wide area networks (WAN). The total latency $L_{i,j}$ of a job $J_{i,j}$ is defined as:
\begin{equation}
 L_{i,j}=D_{i,j}^{comm}+D_{i,j}^{queue}+D_{i,j}^{exec}
\end{equation}
where  $D_{i,j}^{comm}$, $D_{i,j}^{queue}$, and $D_{i,j}^{exec}$ denote communication delay, queuing delay, and execution time, respectively.

Communication delay occurs when a job is offloaded to a remote node. It includes the transmission and propagation delays:
\begin{equation}
  D_{i,j}^{comm}=s_{i,j}^{in} / B_n +d_{prop,n}
\end{equation}                                     
where $s_{i,j}^{in}$  is input data size (bits) of entry task of a job $J_{i,j}$, $B_n$ is available bandwidth (bits/ second) to node $n$, and  $d_{prop,n}$ is propagation delay (seconds).

Jobs assigned to the same VM execute sequentially and may wait in the queue before execution:
\begin{equation}
  D_{i,j}^{queue}=\sum_{J_{i,j} \in Q_{n,v}}D_{i,j}^{exec} 
\end{equation} 
where $Q_{n,v}$ is the set of jobs ahead of job $J_{i,j}$ on VM $v$.

Execution time of a job depends on its constituent task's computational requirement and the node's processing capacity. Let $MI_t$ be the millions of instructions required by task $t$, and $C_n$ the CPU capacity of node $n$. If sufficient CPU cores are available to support DAG parallelism, execution time follows the critical path:
\begin{equation}
   D_{i,j}^{exec} = \max\limits_{b \in B_{i,j}} \sum_{t \in b} \frac{MI_t}{C_n}
\end{equation} 
where $B_{i,j}$ denotes the total set of DAG paths $b$. Otherwise, tasks execute sequentially: 
\begin{equation}
 D_{i,j}^{exec}=\sum_{t \in T_{i,j}} \frac{MI_t}{C_n}
\end{equation} 
where $T_{i,j}$ is the set of tasks in job $J_{i,j}$.

Each job has a deadline $D_{i,j}^{deadline}$ defined by its SLA. A violation occurs when latency exceeds the deadline for that job:
\begin{equation}
SLA_{i,j}=
\begin{cases}
    1 & L_{i,j}>D_{i,j}^{deadline}\\
    0   & \text{otherwise}
\end{cases}
\end{equation} 

During epoch $e$, the total SLA violation rate is defined as:
\begin{equation}
SLA_{rate,e}=\frac{\sum_{J_{i,j} \in J_e} SLA_{i,j}}{\mid{J_e}\mid}
\end{equation}
\subsection{Energy and Carbon Model}
To calculate energy consumption, we model the power usage of each computing node as a function of CPU utilization. The system operates in epochs, and node utilization is sampled every $\Delta$ seconds. At each sampling interval, the CPU utilization of node $n$ is found by aggregating the utilization of all VMs hosted on that node. The power consumption of node $n$ at time $\Delta$ is modeled as:
\begin{equation}
P_n (u_n (\Delta))=((P_n^{idle}+(P_n^{max}-P_n^{idle} ) \times u_n (\Delta)))/3600
\end{equation} 
where $u_n (\Delta)$ represents the CPU utilization rate (0-1) of node $n$.

The energy consumption $EC_n$ (kWh) for a sampling interval is:
\begin{equation}
EC_n (\Delta)=  (P_n (u_n (\Delta)) \times \Delta)/1000
\end{equation} 

The total energy consumed by node $n$ during epoch $e$ is obtained by summing the energy across all sampling intervals: 
\begin{equation}
EC_{n,e}=\sum_{t=1}^TEC_n (\Delta)
\end{equation} 
where $T$ is the number of sampling intervals within an epoch.

To estimate the environmental impact, the total carbon emissions (gCO2) generated by node $n$ during epoch $e$ are computed using the carbon intensity $CI_n$ (gCO2/kWh) of the electricity supply at the corresponding datacenter location:  
\begin{equation}
CO2_{n,e}=EC_{n,e} \times CI_n
\end{equation} 

The total carbon emissions of the edge-fog-cloud system during epoch $e$ are:
\begin{equation}
 CO2_{total,e}= \sum_{n \in N} CO2_{n,e} 
\end{equation} 
where $N$ denotes all computing nodes, including edge devices, fog datacenter servers, and cloud datacenter servers.
\begin{figure}[!t]
  \centering
  \includegraphics[width=\linewidth]{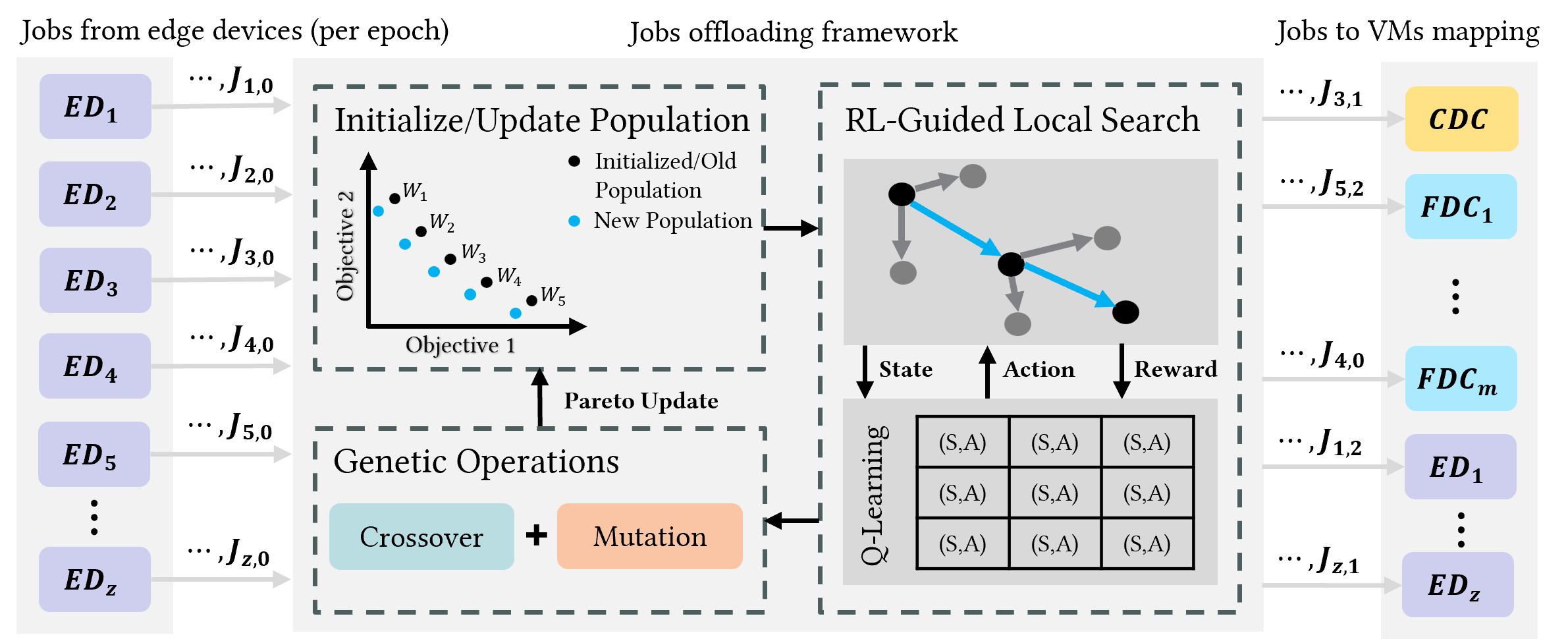}
  \caption{Overview of MERSEM framework.}
  \label{fig2: MERSEM-overview} 
\end{figure}
\section{Problem Formulation}
We consider an edge-fog-cloud computing environment for orchestrating job offloading. The system consists of a central cloud datacenter, geographically distributed fog datacenters, and a set of edge devices that generate computational jobs. Each generated job can be executed at one of the following locations: on the originating edge device, at one of the fog datacenters, or at a cloud datacenter. A job set $J_e$ denotes all the jobs generated in epoch $e$, and $V$ denotes the set of available virtual machines across the infrastructure. Since each VM belongs to a specific node and datacenter, assigning a job to a VM implicitly determines the execution location. We define a binary decision variable:
\begin{equation}
x_{i,j,v}=
\begin{cases}
    1 & \text{if job } \space J_{i,j} \space \text{ is assigned to VM } v\\
    0   & \text{otherwise}
\end{cases}
\end{equation}

The objective is to minimize a weighted combination of the service-level agreement (SLA) violation rate and the total carbon emissions across all epochs:
\begin{equation}
\min \sum_{e=0}^{E} \left( w_s \, SLA_{rate,e} + w_c \, CO2_{total,e} \right)
\end{equation} 
where $SLA_{rate,e}$ represents the SLA violation rate during epoch $e$, and $CO2_{total,e}$ denotes the total carbon emissions generated by the system during epoch $e$. The weighting parameters $w_s$ and $w_c$ control the trade-off between performance and environmental sustainability, and satisfy: 
\begin{equation}
w_s+w_c=1,   w_s,w_c \geq 0 
\end{equation} 

Each job must also be assigned to exactly one VM:
\begin{equation}
\sum_{v \in V}x_{i,j,v}=1,\forall J_{i,j}
\end{equation} 

Multiple jobs can be mapped to the same VM; however, as described earlier, each VM executes assigned jobs sequentially according to the queueing policy.

\section{MERSEM Framework Overview}
Fig. \ref{fig2: MERSEM-overview} illustrates the MERSEM framework for multi-objective job offloading in heterogeneous edge-fog-cloud environments. Jobs genertaed by edge devices are mapped to VMs across edge, fog, and cloud layers to minimize SLA violations and carbon emissions. 

MERSEM integrates RL-guided local search with evolutionary optimization to efficiently explore the solution space. Initially, a population of candidate job-to-VM mappings and a set of weight vectors representing different objective preference are generated. During each generation, a subset of solutions is selected (line 2 of Algorithm 1) and refined using RL-guided local search (line 5) to improve scalarized fitness values. The refined solutions replace the original population members (line 8), after which crossover and mutation operations generate offspring solutions (line 9). A dominance-based update mechanism then evaluates and inserts solutions into the population using fitness and Pareto dominance criteria (line 10). By combining global evolutionary exploration with adaptive RL-based refinement, MERSEM effectively searches the multi-objective space and produces Pareto-optimal scheduling solutions for SLA-focused, carbon-focused, or balanced objectives.
\begin{figure}[!t]
  \centering
  \includegraphics[width=\linewidth]{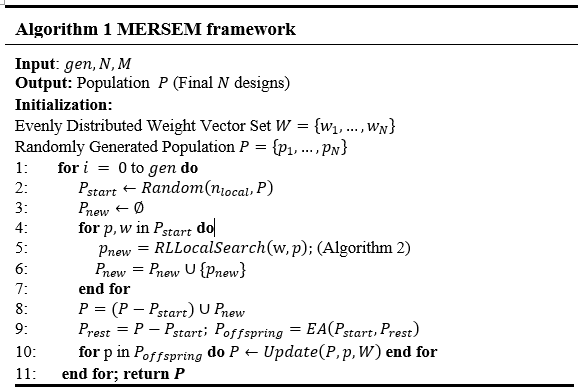}
\end{figure}
\subsection{RL-Guided Local Search}
To improve job offloading solutions, we integrate RL with local search. Our RL agent learns which perturbation strategies are most effective for improving a solution during the search process. Unlike traditional local search that applies random or fixed perturbations, the proposed approach adaptively selects actions based on the current solution state. The RL-guided local search procedure is summarized in Algorithm 2. Given an initial offloading solution $p$ and its associated weight vector $w$, the algorithm refines the solution over multiple episodes (line 3) and perturbation steps (line 9). At each step (lines 10-24), the current solution is encoded into a state, and an action is selected using an $\epsilon{-greedy}$ policy. The action perturbs the solution to generate a new candidate $p^{'}$. The objectives of the perturbed solution are evaluated, and a reward is computed based on fitness improvement and dominance relations. The transition  $(state_{enc},a,r)$ is stored in a trajectory buffer and updates the Q-table at the end of the episode (line 26). 

\textit{State representation}: The RL agent operates on a compact state representation that captures both workload and system performance characteristics. For each candidate solution $state$ (line 4), the state encoding $state_{enc}$  (line 10) is constructed using  scalarized fitness score, time-of-day, normalized SLA violation rate, normalized carbon footprint, number of jobs assigned to each layer (edge, fog, and cloud), and total number of jobs per epoch. To enable efficient state-action mapping, each of these components is discretized into finite bins, allowing the agent to learn over a structured and tractable state space.

\textit{Action space}: The action space contains five perturbation strengths that modify the current solution: $a_1:2\%, a_2:5\%, a_3:8\%, a_4:10\%, \text{ and } a_5:15\%$ job reassignment. For each action, a subset of jobs is randomly reassigned to different nodes (device, fog, or cloud). Smaller perturbations encourage local refinement, while larger ones help escape local optima. 

\textit{Reward function}: The reward reflects both fitness improvement and Pareto dominance, and is defined as:
\begin{equation}
  r=0.7 \times tanh(Obj_{current}-Obj_{new}) + 0.3 \times r_{dom}
\end{equation}
where $Obj_{current}$ is the fitness of the current solution, $Obj_{new}$ is the fitness of the perturbed solution, and $r_{dom}$ is the dominance reward. The dominance reward $r_{dom}$ is defined as:
\begin{equation}
r_{dom}=
\begin{cases}
    1.0 & \text{if } obj_{new}  \text{ dominates } obj_{current} \\
    -0.5 &  \text{if } obj_{current}  \text{ dominates } obj_{new} \\
     0.1 &  \text{otherwise} 
\end{cases}
\end{equation}
The tanh function stabilizes the reward, while the dominance term encourages improvements in the multi-objective space. 

\textit{Trajectory-based learning}: Instead of updating the Q-values after every step, transitions are stored in a trajectory and updated at the end of each episode. The discounted return is computed as $G_t=r_t+ \gamma G_{t+1}$, where $\gamma$ is the discount factor. An additional improvement signal $I=tanh(f_{start}-f_{localBest})$ reflects overall improvements during the trajectory. The learning rate $\alpha$ impacts the updating of the Q-value based on the term $\beta$, defined as: 
\begin{equation}
      \beta=\alpha(G_t+I-Q(state_{enc},a))
\end{equation}
 The full Q-value update is then given by:
 \begin{equation}
        Q(state_{enc},a)=Q(state_{enc},a)+ \beta  
\end{equation}

 \textit{Exploration strategy}: Action selection follows an $\epsilon-greedy$ policy, where a random action is chosen for exploration with probability $\epsilon$; otherwise, the action with the highest Q-value is selected. The exploration rate gradually decays across episodes.
 
Overall, RL-guided local search enables adaptive perturbation of job assignments, balancing local refinement and global exploration to improve job offloading solutions.
\begin{figure}[!t]
  \centering
  \includegraphics[width=\linewidth]{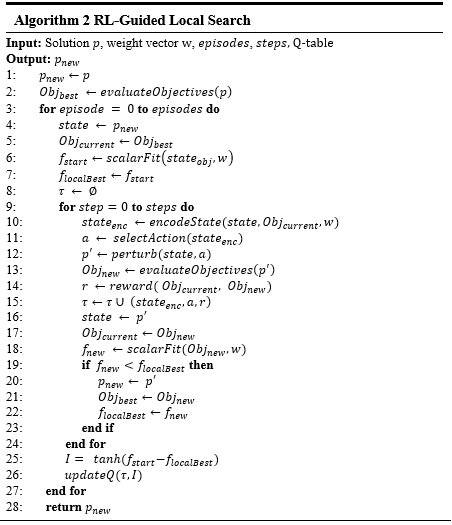}
\end{figure}
\subsection{Evolutionary Algorithm}
While RL-guided local search improves individual solutions, the evolutionary component maintains global exploration and population diversity. As shown in Algorithm 1, crossover and mutation operators are applied to generate offspring solutions $P_{offspring}$ (line 9). Crossover combines job-to-VM assignment patterns from parent solutions, while mutation introduces random variations to preserve diversity. The offspring are then evaluated and the population is updated using a dominance-based replacement strategy (line 10), which considers both scalarized fitness and Pareto dominance.

By combining evolutionary exploration with RL-guided refinement, MERSEM effectively balances exploration and exploitation to identify high-quality offloading solutions that satisfy both performance and sustainability objectives in edge–fog–cloud systems.
\section{Experiments}
\subsection{Experimental Setup}
We evaluate the proposed MERSEM framework against state-of-the-art methods, including Carbon-Aware Grey Wolf Optimization (CAGWO), Crow Search Algorithm (CACSA), and Squirrel Search Algorithm (CASSA) proposed by Pournazari et al. \cite{pournazari2025carbon}, and the hybrid Particle Swarm Optimization and Genetic Algorithm (PSOGA) introduced by Wang et al. \cite{yuping2025pso}, while considering three representative MERSEM solutions derived from the weight formulation in Eq. 15: MERSEM-Carbon (fully carbon-focused), MERSEM-SLA (fully SLA-focused), and MERSEM-Balanced (equal emphasis on carbon and SLA). To ensure fairness in comparison, all frameworks make use of the same population size, time budget, and carbon model.

All experiments are conducted using an extended version of the EdgeCloudSim \cite{sonmez2017edgecloudsim} simulator. The baseline simulated infrastructure consists of one cloud datacenter, four geographically distributed fog datacenters, and 500 edge devices.

The cloud datacenter includes heterogeneous Intel server nodes based on Intel Xeon Platinum 8470, Intel Xeon Gold 6430, and Intel Xeon Silver 4410Y CPUs. Each fog datacenter contains servers with heterogeneous server nodes based on Intel Core Ultra 7, Intel Xeon D-1746TER, and Intel Xeon Gold 6430 CPUs. The nodes differ in CPU cores, clock frequency, power profiles, and memory capacity. The computational capacity ranges from 9,000 to 12,160 MIPS per core. The cloud datacenter hosts 20 servers running 88 virtual machines (VMs), while each fog datacenter includes 5 servers hosting 20 VMs. Each VM is allocated 2-12 CPU cores based on the workload. The system includes 500 edge devices that generate graph workloads. Half of the devices are smartphones with 4 CPU cores (2000 MIPS per core), while the remaining devices are smart camera IoT nodes with 6 CPU cores (2400 MIPS per core). These devices consume 3W idle power and 12W peak power. Network bandwidth between devices and edge datacenters is assumed to be 40 Mbps (WiFi links) and 150 Mbps (MAN link), while fog-to-cloud communication uses 100 Mbps (WAN link) to reflect typical hierarchical network deployments.

The energy consumption of each node is calculated using the power model with a 6-second sampling interval. A centralized orchestrator hosts the algorithmic framework that determines job-to-VM mappings. A workload predictor is assumed based on \cite{qi2026shieldeb} for all frameworks that allows predicting workload for the upcoming epoch (15-minutes). Each comparison framework (including MERSEM) uses the workload predictor at the start of the current epoch and executes its algorithms for the duration of the epoch to generate optimized job-to-VM mappings for the next epoch. The carbon intensity of each datacenter is modeled based on \cite{qi2023shield}, while on-site solar energy generation at fog datacenters and IoT devices is modeled based on \cite{islam2025rl}. All power/energy models, as well as models for latency, bandwidth, compute capability, and carbon intensity are integrated into the simulator discussed earlier.

Key parameters of the MERSEM framework include the number of solutions participating in local search, the number of RL episodes, the number of perturbation steps per episode, and genetic operation rates, which were determined through sensitivity analysis, as discussed in the following section. For the RL-guided local search, the Q-learning agent uses a learning rate  $\alpha=0.15$ and discount factor $\gamma=0.9$. The exploration strategy follows an  $\epsilon-greedy$ policy with an initial exploration rate  $\epsilon=0.4$, which gradually decays during the search process.
\begin{figure}[!h]
  \centering
  \includegraphics[width=\linewidth]{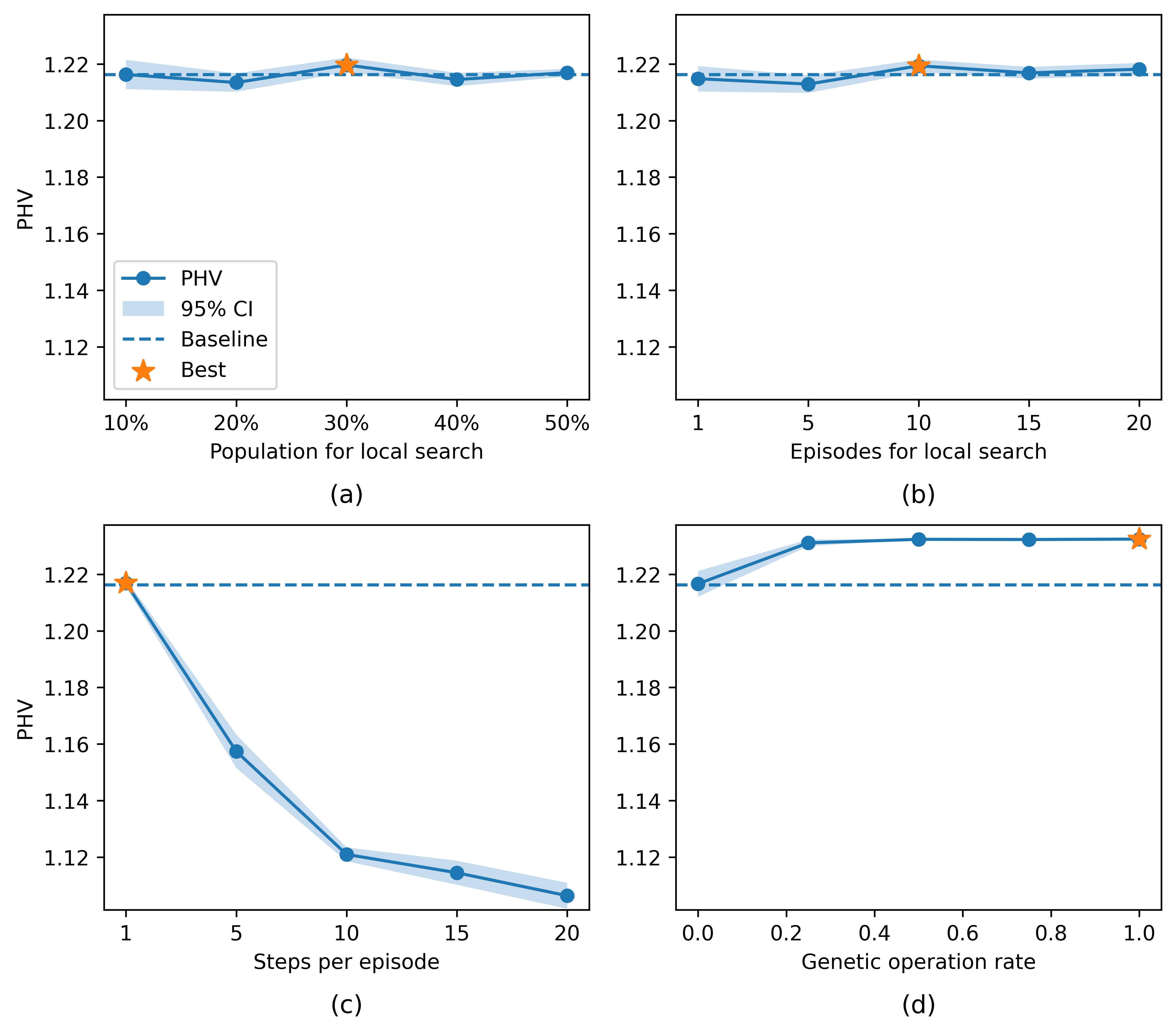}
  \caption{Hyperparameter sensitivity analysis for the MERSEM framework.}
  \label{fig3: sensitivity_analysis}
\end{figure}
\subsection{Hyperparameter Analysis}
We evaluate the sensitivity of the MERSEM algorithm to key hyperparameters by varying one parameter at a time while keeping the others fixed at their baseline values. Performance is measured using the pareto hypervolume (PHV) metric that measures the hypervolume $PHV(P,ref)$ of the solution space that is enclosed by the solution set $P$ and a reference point $ref$. A larger PHV value is indicative of a superior generated solution set by a multi-objective optimization framework in terms of both diversity and quality. The average PHV and $95\%$ confidence intervals are shown in Fig. \ref{fig3: sensitivity_analysis}. It can be observed that PHV remains largely stable when varying the local search population and number of episodes, indicating low sensitivity to these parameters. Increasing the number of steps per episode reduces PHV due to higher computational overhead, which limits the number of evolutionary iterations. The genetic operation rate shows stable performance, with slightly improved PHV near 1.0 due to increased exploration.Based on this analysis, the configuration of 30$\%$ local-search participation, 10 episodes, 1 step per episode, and a genetic operation rate of 1.0 is selected to balance exploration and exploitation while maintaining a high-quality Pareto solution set.
\begin{figure}
  \centering
  \includegraphics[width=\linewidth]{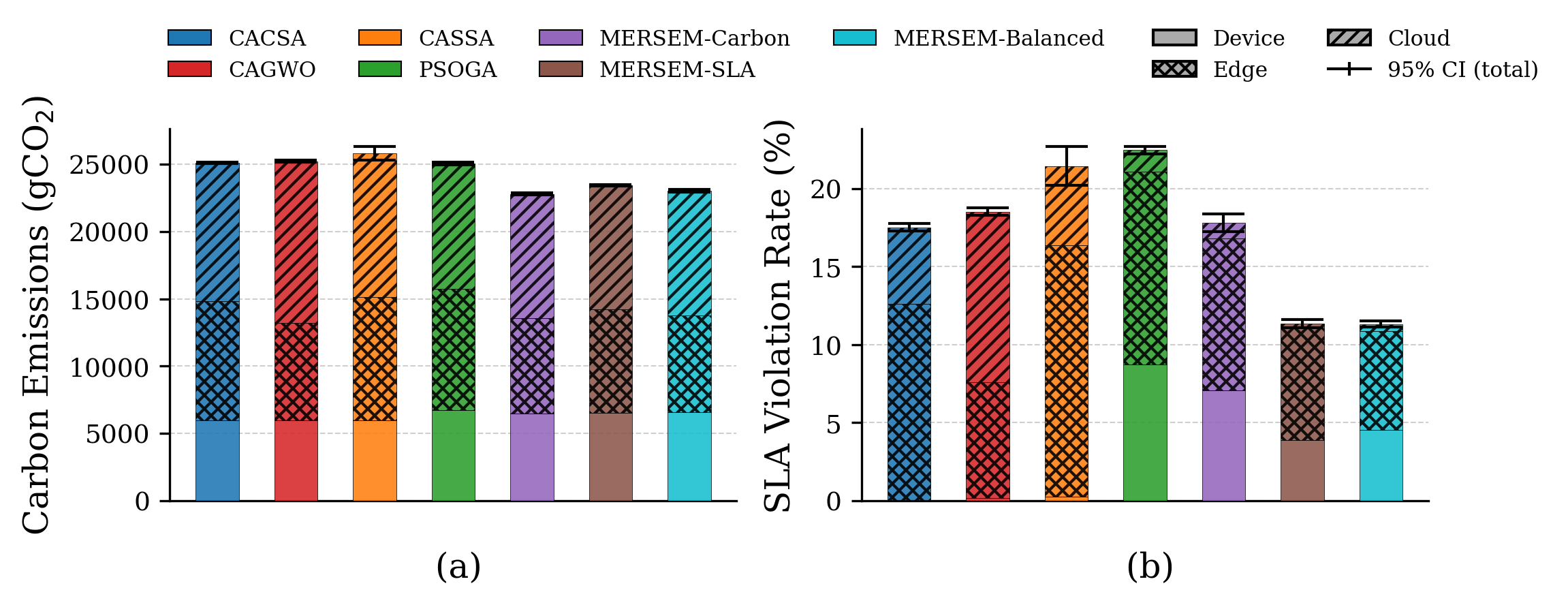}
  \caption{Comparison of (a) Carbon Emissions and (b) SLA Violation Rate across frameworks.}
  \label{fig4:main_results}
\end{figure}
\subsection{State-of-the-art Comparison }
Fig. \ref{fig4:main_results} compares MERSEM with state-of-the-art frameworks in terms of carbon emissions and SLA violation rate, where stacked bars show contributions from device, fog, and cloud layers, and error bars indicate the 95$\%$ confidence interval across runs. 

In Fig. \ref{fig4:main_results}(a), it can be observed that the state-of-the-art methods (CACSA, CAGWO, CASSA, and PSOGA) generate approximately 25,000-26,000 gCO2 emissions. In contrast, MERSEM-Carbon and MERSEM-Balanced reduce emissions to about 23,000 gCO2, achieving 10-12$\%$ reduction. MERSEM-SLA produces slightly higher emissions around 23,500 gCO2 due to its focus on performance but still achieves 8-10$\%$ reductions compared to state-of-the-art approaches. 

Fig. \ref{fig4:main_results}(b) shows that state-of-the-art methods exhibit 17-23$\%$ SLA violation rates. MERSEM-SLA achieves the lowest violation rate 11-12$\%$, representing a 35-45$\%$ reduction compared to the worst state-of-the-art. MERSEM-Balanced achieves a similar  violation rate while maintaining low emissions, whereas MERSEM-Carbon shows slightly higher violations 18$\%$ due to its emphasis on carbon minimization. 

Overall, MERSEM significantly improves performance, reducing carbon emissions by up to 12$\%$ and SLA violations by up to 45$\%$ compared to existing methods.
\begin{figure}[!t]
  \centering
  \includegraphics[width=\linewidth]{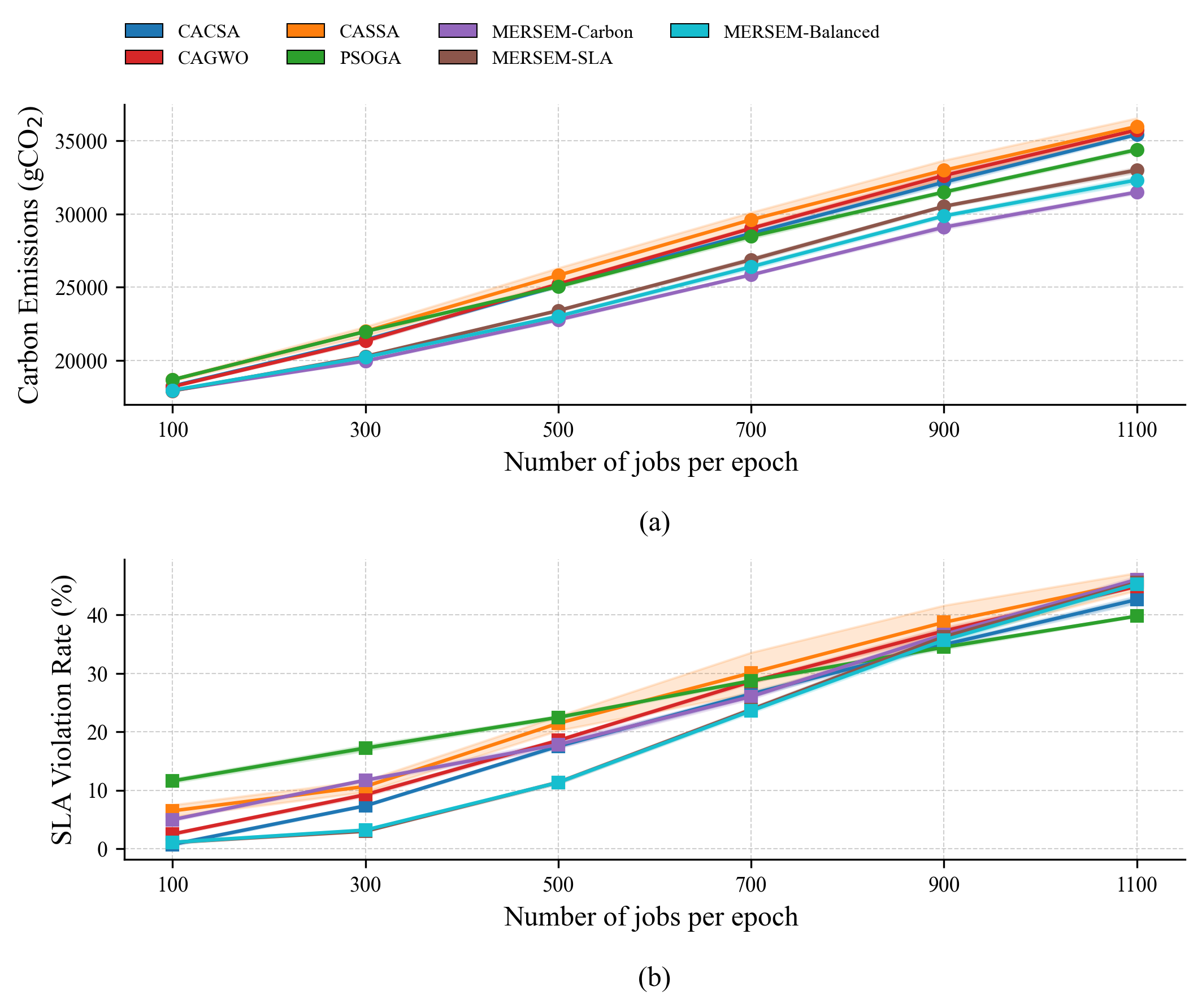}
  \caption{Comparison of (a) Carbon Emissions and (b) SLA Violation Rate across frameworks as the number of jobs per epoch scales up from 100 to 1100 jobs.}
  \label{fig5:sclability_analysis_1}
\end{figure}
\subsection{Scalability Analysis}
Fig. \ref{fig5:sclability_analysis_1} evaluates the scalability of all frameworks by increasing the number of jobs per epoch from 100 to 1100, while keeping the number of devices, fog, and cloud datacenters constant.

Fig. \ref{fig5:sclability_analysis_1}(a) shows carbon emissions as workload increases. At 100 jobs, all methods produce similar emissions 18000-19000 gCO2. As workload grows, differences become clearer. At 500 jobs, state-of-the-art methods produce about 25000-26000 gCO2, while MERSEM-Carbon and MERSEM-Balanced maintain lower emissions 23,000-24,000 gCO2. At 1100 jobs, state-of-the-art methods reach 35000-36000 gCO2, whereas MERSEM-Carbon limits emissions to about 32000 gCO2, achieving roughly 8-10$\%$ lower emissions, and MERSEM-Balanced maintains around 33,000 gCO2. MERSEM-SLA produces slightly higher emissions due to its focus on service performance.

Fig. \ref{fig5:sclability_analysis_1}(b) shows SLA violation rates as workloads increase. At 100 jobs, violations remain 0-10$\%$ for all methods. At 500 jobs, state-of-the-art methods rise to 20-25$\%$, while MERSEM-Balanced maintains 15-18$\%$, and MERSEM-SLA performs similarly or slightly better. At 1100 jobs, state-of-the-art methods reach 40-45$\%$ violations, whereas MERSEM variants remain lower at 40-43$\%$, indicating improved service reliability under heavy workloads. Overall, the results show that MERSEM maintains lower carbon emissions and competitive SLA performance as workload increases, demonstrating its scalability for large-scale edge-cloud job offloading environments.
\begin{figure}[!h]
  \centering
  \includegraphics[width=\linewidth]{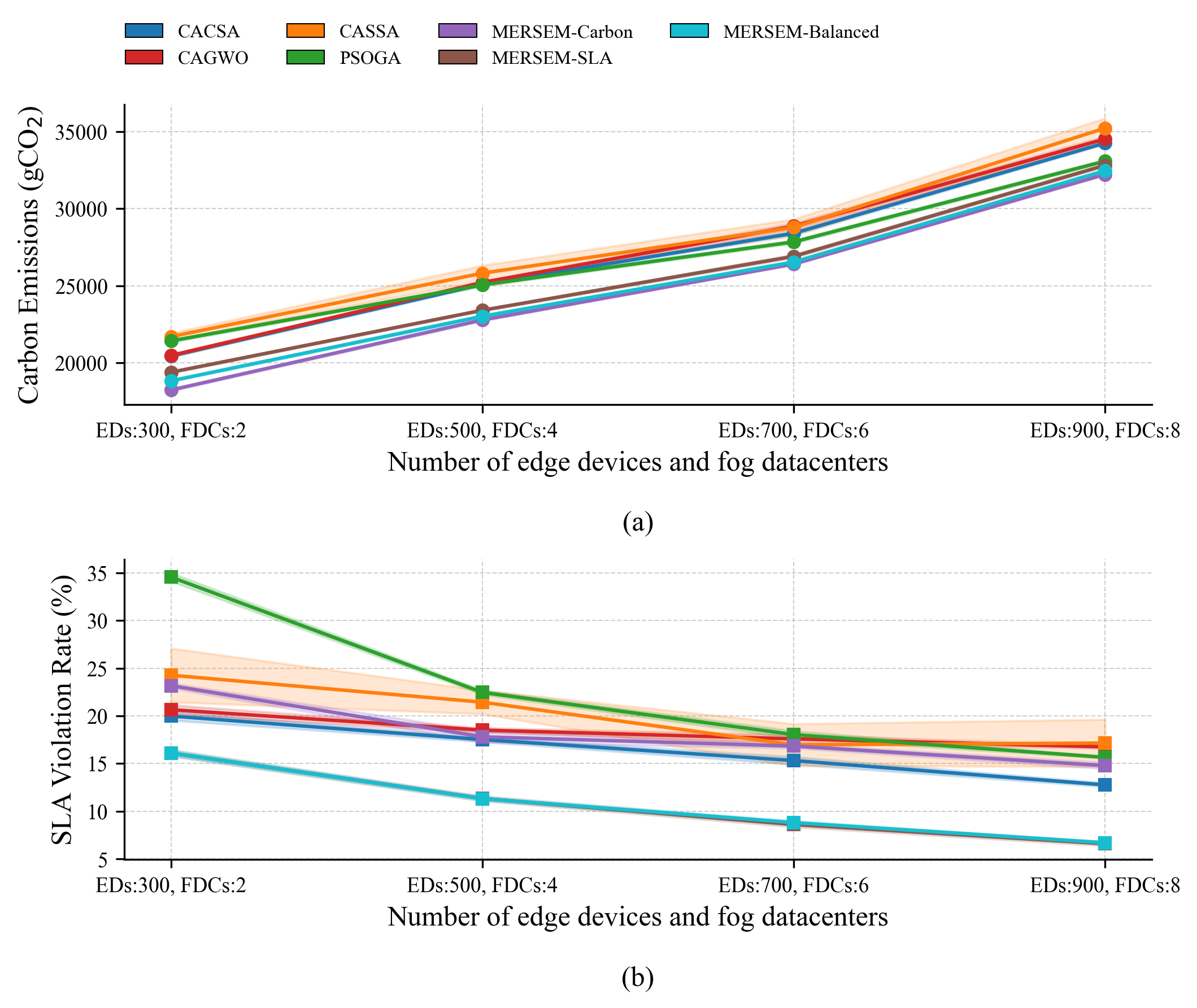}
  \caption{Comparison of (a) Carbon Emissions and (b) SLA Violation Rate across frameworks as the number of Edge devices (300-900) and Fog datacenters (2-8) scale up.}
  \label{fig6:scalability_analysis_2}
\end{figure}

To evaluate the impact of scalability of compute resources across frameworks, we keep the workload fixed at 500 jobs per epoch, and scale up the number of edge devices and fog datacenters, with results shown in Fig. \ref{fig6:scalability_analysis_2}.

From Fig. \ref{fig6:scalability_analysis_2}(a), we can see that as the infrastructure scales up, carbon emissions increase accordingly due to greater resource availability and distributed execution. At 300 EDs and 2 FDCs, state-of-the-art methods produce about 20,500-21,800 gCO2, while MERSEM-Carbon and MERSEM-Balanced achieve lower emissions of around 18,200-18,800 gCO2. As the system scales up to 900 EDs and 8 FDCs, state-of-the-art emissions increase to 33,200-35,200 gCO2, while MERSEM-Carbon limits emissions to 32,000 gCO2. At the same time, MERSEM-Balanced and MERSEM-SLA remain slightly higher in carbon emissions than MERSEM-Carbon, but still lower than the state-of-the-art methods. 

Fig. \ref{fig6:scalability_analysis_2}(b) shows the SLA violation rates of frameworks at different system scales. At 300 EDs and 2 FDCs, state-of-the-art methods exhibit 20-24$\%$ SLA violation rates, whereas MERSEM-SLA achieves 16$\%$. As the infrastructure scales up to 900 EDs and 8 FDCs, the state-of-the-art methods' violation rate decreases to 13-17$\%$, whereas MERSEM variants achieve the lowest rates, at 6-7$\%$. Overall, the results from Fig. \ref{fig6:scalability_analysis_2} show that MERSEM maintains lower carbon emissions and significantly reduces SLA violations as the edge-cloud system scales up, demonstrating improved efficiency and service reliability.
\begin{figure}[!h]
  \centering
  \includegraphics[width=\linewidth]{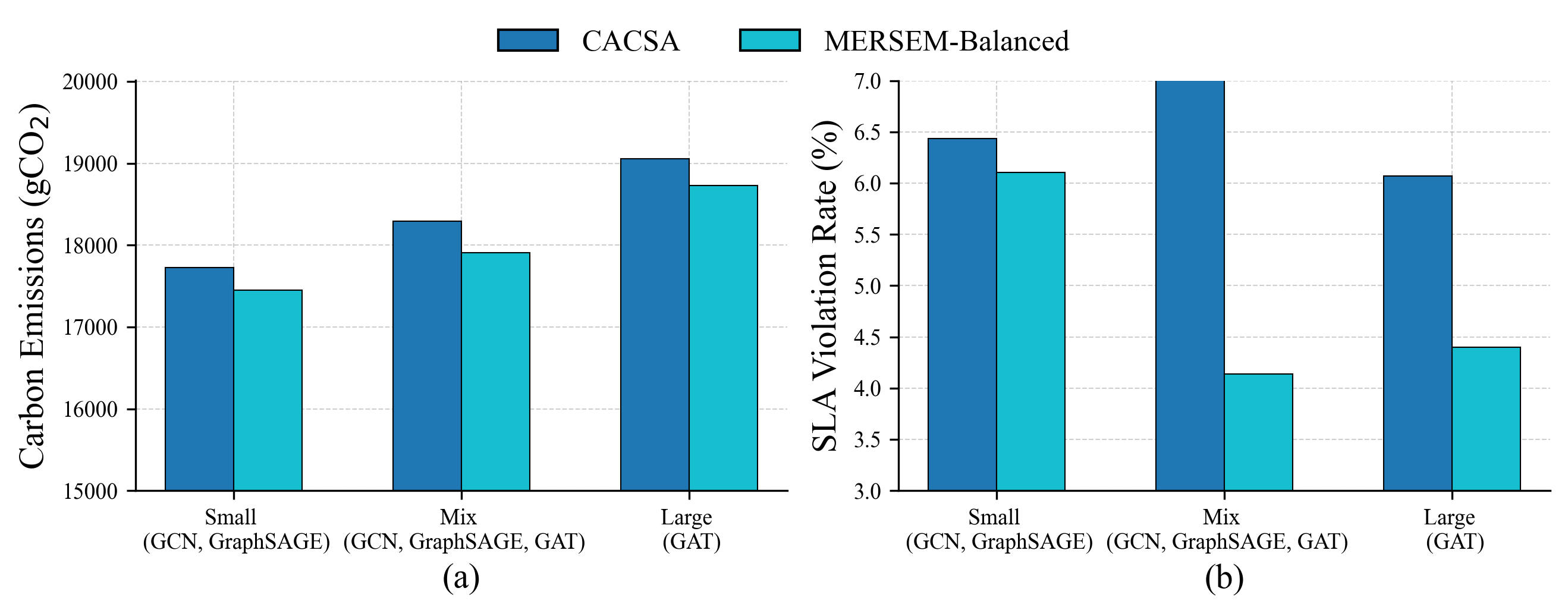}
  \caption{Comparison of (a) Carbon Emissions and (b) SLA Violation Rate between MERSEM and best state-of-the-art, for GNN-based graph analytics workloads.}
  \label{fig7:analysis_on_GNN_workloads}
\end{figure}
\subsection{Performance Analysis on GNN Workloads}
To evaluate the robustness of MERSEM under different GNN-based graph analytics workloads, we conducted experiments using three datasets and models: the CORA dataset with GraphSAGE, the Amazon dataset with GCN, and the CiteSeer dataset with GAT. Due to its significantly higher computational complexity, GAT workloads are categorized as large jobs, while GCN and GraphSAGE are treated as small jobs. Additionally, a mixed workload scenario is constructed by combining jobs from all three models. Each scheduling epoch consists of 500 jobs.

Fig. \ref{fig7:analysis_on_GNN_workloads} presents a comparison between MERSEM-Balanced and the best state-of-the-art approach (CACSA) in terms of carbon emissions and SLA violation rate across small, mixed, and large workloads. As expected, carbon emissions increase with workload intensity, with large (GAT) workloads exhibiting the highest emissions, followed by mixed and small workloads. Across all workload types, MERSEM-Balanced consistently achieves lower carbon emissions than CACSA, demonstrating improved carbon efficiency. In terms of SLA violations, CACSA shows relatively high violation rates, particularly in the mixed workload scenario. In contrast, MERSEM-Balanced significantly reduces SLA violations for mixed and large workloads, indicating better adaptability to heterogeneous and compute-intensive job distributions. Although the improvement is more modest for small workloads, MERSEM-Balanced still maintains lower violation rates compared to CACSA. Overall, these results highlight MERSEM-Balanced's effectiveness in jointly optimizing carbon efficiency and service reliability, especially under complex and high-demand GNN workload conditions.
\section{Conclusion}
This paper presented MERSEM, a hybrid evolutionary reinforcement learning framework for sustainable graph workload scheduling in heterogeneous edge-cloud environments. MERSEM addresses a key gap in carbon-aware resource management by jointly optimizing application-level SLA compliance and operational carbon emissions, while explicitly modeling the DAG structure, limited parallelism, and communication sensitivity of graph analytics workloads. By combining population-based evolutionary search with RL-guided local refinement, MERSEM effectively balances global exploration and adaptive exploitation under dynamic system conditions. Compared to the best state-of-the-art approach, MERSEM achieves up to 45$\%$ reductions in SLA violations and 12$\%$ reductions in carbon emissions, while maintaining strong scalability as workload intensity, device count, and fog infrastructure size increase. Moreover, evaluation on small, large, and mixed GNN workloads confirms MERSEM’s robustness and ability to sustain a favorable performance-sustainability trade-off under computation-intensive scenarios. By enabling flexible configurations that prioritize SLA performance, carbon efficiency, or balanced objectives, MERSEM provides a practical foundation for next-generation sustainable edge-cloud orchestration.
\begin{acks}
This research was made possible with support from HPE and grants from the National Science Foundation (CCF-2324514, CNS-2132385)
\end{acks}
\bibliographystyle{ACM-Reference-Format}
\bibliography{sample-base}
\end{document}